\documentclass[aps,prd,reprint,superscriptaddress]{revtex4-1}
\usepackage{hyperref}
\usepackage{graphicx}
\usepackage{enumerate}
\hypersetup{
    pdfnewwindow=true,      % links in new window
    colorlinks=true,        % false: boxed links; true: colored links
    linkcolor=blue,         % color of internal links
    citecolor=black,        % color of links to bibliography
    filecolor=blue,         % color of file links
    urlcolor=blue           % color of external links
}

\usepackage{subfigure}
\usepackage{amsmath}
\usepackage{amssymb}
\usepackage{amsfonts}
\RequirePackage{lineno}
\usepackage{color}
\usepackage[normalem]{ulem}
\usepackage{soul}
\usepackage{lineno}
\usepackage[table]{xcolor}

\def\be{\begin{equation}}
\def\ee{\end{equation}}

\providecommand{\araa}[0]{Ann.\ Rev. Astron. Astroph. }
\providecommand{\aap}[0]{Astron. Astrophys. }

\providecommand{\astropart}[0]{Astropart. Phys. }
\providecommand{\apj}[0]{Astrophys. J.~}
\providecommand{\apjl}[0]{Astrophys. J. Lett. }

\providecommand{\cqg}[0]{CQG~} % Classical and Quantum Gravity
\providecommand{\jcap}[0]{JCAP }
\providecommand{\ji}[0]{J. Instrum. } % Journal of Instrumentation

\providecommand{\nature}[0]{Nature }
\providecommand{\nimA}[0]{Nucl. Instrum. Meth. A } % Nuclear Instruments and Methods in Physics Research A

\providecommand{\prd}{Phys. Rev. D. }
\providecommand{\prl}[0]{Phys. Rev. Lett. }
\providecommand{\rpp}[0]{Rep. Prog. Phys. } % Reports on Progress in Physics
\providecommand{\rmp}{Rev. Mod. Phys. } % Reviews of Modern Physics

\providecommand{\epjc}[0]{Eur. Phys. J. C } % European Physical Journal C

\newcommand{\PESKYNINTY}{{\ensuremath{590\,\mathrm{deg^2}}}} % LALInference: 90% credible region
\newcommand{\PESKYFIFTY}{{\ensuremath{140\,\mathrm{deg^2}}}} % LALInference: 50% credible region
\newcommand{\MONESCOMPACT}{{\ensuremath{36_{-4}^{+5}}}} % PE:Overall source (physical) mass m1
\newcommand{\MTWOSCOMPACT}{{\ensuremath{29_{-4}^{+4}}}} % PE: Overall source (physical) mass m2
\newcommand{\DISTANCECOMPACT}{{\ensuremath{410_{-180}^{+160}}}} % PE: Overall distance in Mpc
\newcommand{\CBCEVENTIFAR}{{\ensuremath{203\,000}}}
\newcommand{\CBCEVENTSIGMA}{{\ensuremath{5.1}}}
\newcommand{\REDSHIFTCOMPACT}{{\ensuremath{0.09_{-0.04}^{+0.03}}}} % PE: Overall redshift
\newcommand{\CHIRPDURATION}{{\ensuremath{0.2}}} % s
\newcommand{\ERADERGSIMPLE}{{\ensuremath{5 \times 10^{54}}}} % PE: Overall radiated energy in ergs
\begin{document}

\title{High-energy Neutrino follow-up search of Gravitational Wave Event \\GW150914 with ANTARES and IceCube}

%%%%%%%%%%%%%%%%%%%%%%%%%%%%%%%%%%%%%%%%%%%%%%%%%%%%%%%%%%%%%%%%%%%%%%

\begin{abstract}
We present the high-energy-neutrino follow-up observations of the first gravitational wave transient GW150914 observed by the Advanced LIGO detectors on Sept.~14$^{\rm th}$, 2015. We search for coincident neutrino candidates within the data recorded by the IceCube and \textsc{Antares} neutrino detectors. A possible joint detection could be used in targeted electromagnetic follow-up observations, given the significantly better angular resolution of neutrino events compared to gravitational waves. We find no neutrino candidates in both temporal and spatial coincidence with the gravitational wave event. Within $\pm 500$\,s of the gravitational wave event, the number of neutrino candidates detected by IceCube and \textsc{Antares} were three and zero, respectively. This is consistent with the expected atmospheric background, and none of the neutrino candidates were directionally coincident with GW150914. We use this non-detection to constrain neutrino emission from the gravitational-wave event.
\end{abstract}

\keywords{}

% put authors to end of paper
  \let\mymaketitle\maketitle
  \let\myauthor\author
  \let\myaffiliation\affiliation
  \author{ANTARES Collaboration}
  \author{IceCube Collaboration}
  \author{LIGO Scientific Collaboration}
  \author{Virgo Collaboration}
  \thanks{Full author list given at the end of the article.}

% put authors to beginning
%  \include{ANTARES_authorlist}
%  \include{ICECUBE_authorlist}
%  \include{LVC_authorlist_new}

\maketitle

\section{Introduction}

Advanced LIGO's first observation periods \cite{2015CQGra..32g4001T,2013arXiv1304.0670T} represent a major step in probing the dynamical origin of high-energy emission from cosmic transients \cite{2013CQGra..30l3001B}. The significant improvement in gravitational wave (GW) search sensitivity enables a comprehensive multimessenger observational effort involving partner electromagnetic observatories from radio to gamma-rays, as well as neutrino detectors. The goals of multimessenger observations are to gain a more complete understanding of cosmic processes through a combination of information from different probes, and to increase search sensitivity over an analysis using a single messenger \cite{2013APh....45...56S,2013RvMP...85.1401A,2014ApJ...795...43F}.

The merger of neutron stars and black holes, and potentially massive stellar core collapse with rapidly rotating cores, are expected to be significant sources of GWs \cite{2013CQGra..30l3001B}. These events can result in a black hole plus accretion disk system that drives a relativistic outflow \cite{1989Natur.340..126E,1993ApJ...405..273W}. Energy dissipation in the outflow produces non-thermal, high-energy radiation that is observed as gamma-ray bursts (GRBs), and may have a $\gg$\,GeV neutrino component at comparable luminosities. %Joint GW and neutrino observations could probe the connection between the physics of relativistic outflow and the dynamics and formation of the progenitor that drives the outflow.

Multiple detectors have been built that can search for this high-energy neutrino signature, including the IceCube Neutrino Observatory\textemdash a cubic-kilometer facility at the South Pole \cite{2006APh....26..155I,2009NIMPA.601..294A,2010NIMPA.618..139A}, and \textsc{Antares} \cite{ANTARES,2007NIMPA.570..107A,2005NIMPA.555..132A} in the Mediterranean sea. The construction of the KM3NeT cubic-kilometer scale neutrino detector in the Mediterranean Sea has started in December 2015 with the  successful deployment of the first detection string \cite{2016arXiv160107459A}. IceCube is planning a substantial increase in sensitivity with near-future upgrades \cite{2014arXiv1412.5106I,2014arXiv1401.2046T}. Another facility, the Baikal Neutrino Telescope is also planning an upgrade to cubic-kilometer volume \cite{Avrorin2011S13}. An astrophysical high-energy neutrino flux has recently been discovered by IceCube \cite{2014PhRvL.113j1101A,2015PhRvL.115h1102A,2015PhRvD..91b2001A,2015ApJ...809...98A}, demonstrating the production of non-thermal high-energy neutrinos. The specific origin of this neutrino flux is currently unknown. Multimessenger analyses constraining the common sources of high-energy neutrinos and GWs have been carried out in the past with both \textsc{Antares} and IceCube \cite{2011PhRvL.107y1101B,2013JCAP...06..008A,2014PhRvD..90j2002A}.

On Sept. 14$^{\rm th}$, 2015 at 09:50:45~UTC, a highly significant GW signal was recorded by the LIGO Hanford, WA and Livingston, LA detectors \cite{G184098}. The event, labeled GW150914, was produced by a stellar-mass binary black hole merger at redshift $z= \REDSHIFTCOMPACT$. The reconstructed mass of each black hole is $\sim 30$\,M$_\odot$. Such a system may produce electromagnetic emission and emit neutrinos if the merger happens in a sufficiently baryon-dense environment, and a black hole plus accretion disk system is formed \cite{2002ARA&A..40..137M}. Current consensus is that such a scenario is unlikely, nevertheless, there are no significant observational constraints.

Here we report the results of a neutrino follow-up search of GW150914 using \textsc{Antares} and IceCube. After brief descriptions of the GW search (Section \ref{sec:GWanalysis}) and the neutrino follow-up (Section \ref{sec:neutrinoanalysis}), we present the joint analysis, results of the search and source constraints, and conclusions (Section \ref{sec:results}).

\section{Gravitational Wave Data Analysis and Discovery}
\label{sec:GWanalysis}

%Before the first Advanced LIGO observation run (O1) began on Sep.~18$^{\rm th}$,~2015 at 15:00:00~UTC \cite{2013arXiv1304.0670T}, LIGO conducted a series of ``engineering runs" in order to test operation, data collection and analysis. The final engineering run (ER8) began on Aug.~17$^{\rm th}$,~2015, 15:00:00~UTC, and lasted until the start of O1. Critical software were frozen by Aug.~30$^{\rm th}$,~2015. Calibration was complete by Sept.~12$^{\rm th}$,~2015.

GW150914 was initially identified by low-latency searches for generic GW transients \cite{2008CQGra..25k4029K,CWB2G:2015all,Burstcompanion}. %The cWB pipeline performs a coherent, constrained maximum likelihood analysis in order to detect and reconstruct GW transients, using minimal assumptions about gravitational waveform morphology.
Subsequent analysis with three independent matched-filter analyses using models of compact binary coalescence waveforms \cite{CBCcompanion,PEcompanion} confirmed that the event was produced by the merger of two black holes. The analyses established a false alarm rate of less than 1 event per \CBCEVENTIFAR~years, equivalent to a significance $>\CBCEVENTSIGMA\,\sigma$ \cite{G184098}. Source parameters were reconstructed using the LALInference package \cite{2013PhRvD..88f2001A,2015PhRvD..91d2003V,PEcompanion}, finding black-hole masses \MONESCOMPACT\,M$_\odot$ and \MTWOSCOMPACT\,M$_{\odot}$ and luminosity distance $D_{\rm gw} = \DISTANCECOMPACT$\,Mpc, where the error ranges correspond to the range of the 90\% credible interval. The duration of the signal within LIGO's sensitive band was $\CHIRPDURATION$\,s.

The directional point spread function (sky map) of the GW event was computed through the full parameter estimation of the signal, carried out using the LALInference package \cite{2013PhRvD..88f2001A,2015PhRvD..91d2003V}. The LALInference results presented here account for calibration uncertainty in the GW strain signal. The sky map is shown in Fig. \ref{figure:skymap}. At 90\% (50\%) credible level (CL), the sky map covers \PESKYNINTY~(\PESKYFIFTY).

%. cWB reconstructs the signal likelihood as a function of source direction over the sky (an updated version of \cite{2008CQGra..25k4029K,CWB2G:2015all}). The sky map is shown in Fig. \ref{figure:skymap}. At 90\% (50\%) credible level, the sky map covers 308$^\circ$ (98$^\circ$).

\section{High-energy Neutrino Coincidence Search}
\label{sec:neutrinoanalysis}

High-energy neutrino observatories are primarily sensitive to neutrinos with $\gg\,$GeV energies. IceCube and \textsc{Antares} are both sensitive to through-going muons (called track events), produced by neutrinos near the detector, above \text{$\sim 100\,$GeV}. In this analysis, \textsc{Antares} data include only up-going tracks for events originating from the Southern hemisphere, while IceCube data include both up-going tracks (from the Northern hemisphere) as well as down-going tracks (from the Southern hemisphere). The energy threshold of neutrino candidates increases in the Southern hemisphere for IceCube, since downward-going atmospheric muons are not filtered by the Earth, greatly increasing the background at lower energies. Neutrino times of arrival are determined at $\mu$s precision.

Since neutrino telescopes continuously take data observing the whole sky, it is possible to look back and search for neutrino counterparts to an interesting GW signal at any time around the GW observation. %Therefore, correlations on essentially any time scale and time differences can be studied.

To search for neutrinos coincident with GW150914, we used a time window of $\pm 500$\,s around the GW transient. This search window, which was used in previous GW-neutrino searches, is a conservative, observation-based upper limit on the plausible emission of GWs and high-energy neutrinos in the case of GRBs, which are thought to be driven by a stellar-mass black hole---accretion disk system \cite{2011APh....35....1B}. While the relative time of arrival of GWs and neutrinos can be informative \cite{2003PhRvD..68h3001R,2012PhRvD..86h3007B,2014PhRvD..90j1301B}, here we do not use detailed temporal information beyond the $\pm 500$\,s time window.

The search for high-energy neutrino candidates recorded by IceCube within $\pm500\,$s of GW150914 used IceCube's online event stream. The online event stream implements an event selection similar to the event selection used for neutrino point source searches \cite{2014ApJ...796..109A}, but optimized for real-time performance at the South Pole.  This event selection consists primarily of cosmic-ray-induced background events, with an expectation per 1000 seconds of 2.2 events in the Northern sky (atmospheric neutrinos), and 2.2 events in the Southern sky (high-energy atmospheric muons).  In the search window of $\pm 500$\,s centered on the GW alert time (see below), one event was found in the Southern sky and two in the Northern sky, which is consistent with the background expectation. The properties of these events are listed in Table~\ref{table:neutrinos}. The neutrino candidates' directions are shown in Fig. \ref{figure:skymap}.

The muon energy in Table~\ref{table:neutrinos} is reconstructed assuming a single muon is producing the event. While the event from the Southern hemisphere has a significantly greater reconstructed energy \cite{2014JInst...9P3009A} than the other two events, $12.5\%$ of the background events in the same declination range in the Southern hemisphere have energies in excess of the one observed.  The intense flux of atmospheric muons and bundles of muons that constitute the background for IceCube in the Southern hemisphere gradually falls as the cosmic ray flux declines with energy \cite{2015arXiv150607981I}. The use of energy cuts to remove most of this background is the reason that IceCube's sensitivity in the Southern sky is shifted to higher energies.

An additional search was performed using the high-energy starting event selection described in \cite{2014PhRvL.113j1101A}. No events were found in coincidence with GW150914.

The IceCube detector also has sensitivity to outbursts of MeV neutrinos (as occur for example in core-collapse supernovae) via a sudden increase in the photomultiplier rates \cite{2011A&A...535A.109A}.  The global photomultiplier noise rate is monitored continuously, and deviations sufficient to trigger the lowest-level of alert occur roughly once per hour.  No alert was triggered during the $\pm 500$ second time-window around the GW candidate event.

\begin{table}
\begin{tabular}{c|c|c|c|c|c|c}
  \hline
  $\#$ & $\Delta T$ [s] & RA [h] & Dec [$^\circ$] & $\sigma_{\mu}^{\rm rec}$ [$^\circ$] & E$_{\mu}^{\rm rec}$ [TeV] & fraction \\ \hline
  1    &   $+37.2$  &  8.84         & $-16.6$   & $0.35$  & 175     & 12.5\%\\
  2    &  $+163.2$  & 11.13         &  12.0     & $1.95$  &   1.22  & 26.5\%\\
  3    &  $+311.4$  & $-7.23$       &   8.4     & $0.47$  &   0.33  & 98.4\%\\
  \hline
\end{tabular}
  \caption{Parameters of neutrino candidates identified by IceCube within the $\pm 500$\,s time window around GW150914. $\Delta T$ is the time of arrival of the neutrino candidates relative to that of GW150914. E$_{\mu}^{\rm rec}$ is the reconstructed muon energy. $\sigma_{\mu}^{\rm rec}$ is the angular uncertainty of the reconstructed track direction \cite{2014PhRvD..89j2004A}. The last column shows the fraction of background neutrino candidates with higher reconstructed energy at the same declination ($\pm 5^{\circ}$).}
  \label{table:neutrinos}
\end{table}
% RA in deg
% 132.6
% 167.0
% 251.5

The search for coincident neutrinos for \textsc{Antares} within $\pm 500\,$s of GW150914 used \textsc{Antares}'s online reconstruction pipeline \cite{2015arXiv150801180A}. A fast and robust algorithm \cite{2011APh....34..652A} selected up-going neutrino candidates with $\sim$\,mHz rate, with atmospheric muon contamination less than $10\%$. In addition, to reduce the background of atmospheric neutrinos \cite{2013EPJC...73.2606A}, a requirement of a minimum reconstructed energy reduced the online event rate to 1.2\,events/day. Consequently, for \textsc{Antares} the expected number of neutrino candidates from the Southern sky in a 1000\,s window in the Southern sky is 0.014. We found no neutrino events from \textsc{Antares} that were temporally coincident with GW150914. This is consistent with the expected background event rate.

\section{Results}
\label{sec:results}

\subsection{Joint analysis}

\begin{figure}
\begin{center}
\resizebox{0.49\textwidth}{!}{\includegraphics{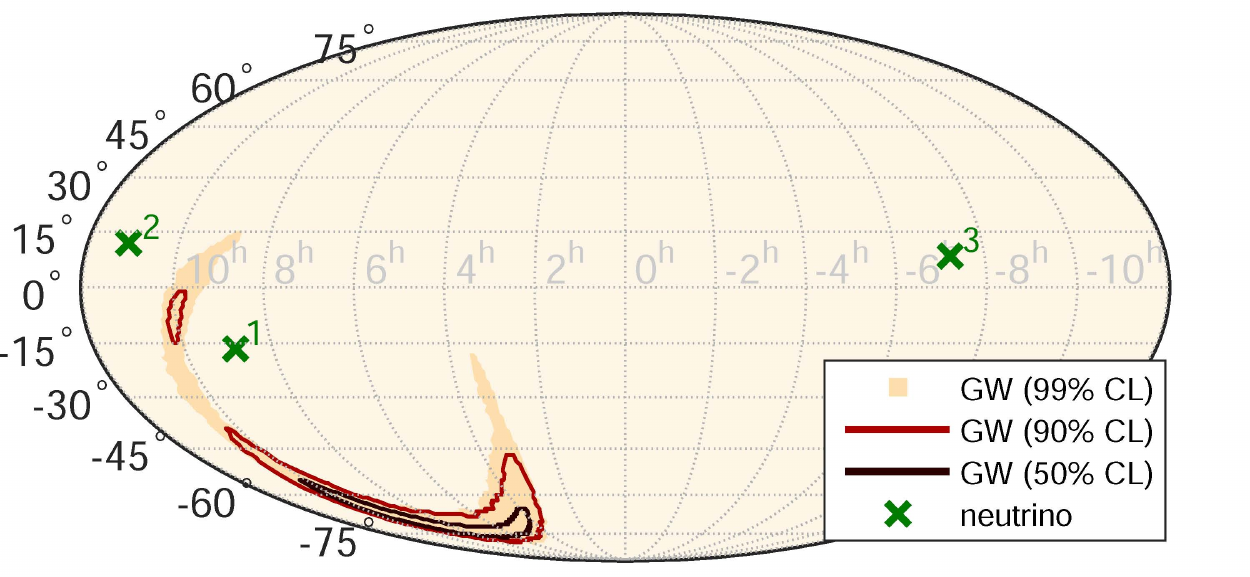}}
\end{center}
\caption{GW skymap in equatorial coordinates, showing the reconstructed probability density contours of the GW event at 50\%, 90\% and 99\% CL, and the reconstructed directions of high-energy neutrino candidates detected by IceCube (crosses) during a $\pm 500$\,s time window around the GW event. The neutrino directional uncertainties are $<1^\circ$ and are not shown. GW shading indicates the reconstructed probability density of the GW event, darker regions corresponding to higher probability. Neutrino numbers refer to the first column of Table~\ref{table:neutrinos}.}
\label{figure:skymap}
\end{figure}

We carried out the joint GW and neutrino search following the analysis developed for previous GW and neutrino datasets using initial GW detectors \cite{2012PhRvD..85j3004B,2014PhRvD..90j2002A,2011PhRvL.107y1101B,2011APh....35....1B}. After identifying the GW event GW150914 with the cWB pipeline, we used reconstructed neutrino candidates to search for temporal and directional coincidences between GW150914 and neutrinos. We assumed that the \emph{a priori} source directional distribution is uniform. For temporal coincidence, we searched within a $\pm 500$\,s time window around GW150914.

%The relative propagation time for $\gg$\,GeV neutrinos over the distance to the source, compared to that of light, is expected to be $\ll1\,$s. The relative propagation time between neutrinos and GWs may change, however, in alternative gravity models . This could be probed by a joint GW-neutrino detection by comparing the times of arrivals with the expected emission time frame.

The relative difference in propagation time for $\gg$\,GeV neutrinos and GWs (which travel at the speed of light in general relativity) traveling to
Earth from the source is expected to be $\ll1\,$s.  The relative propagation time between neutrinos and GWs may change in alternative gravity models \cite{2007NatPh...3...87J,GRcompanion}. However, discrepancies from general relativity could in principle be probed with a joint GW-neutrino detection by comparing the arrival times against the expected time frame of emission.

Directionally, we searched for overlap between the GW sky map and the neutrino point spread functions, assumed to be Gaussian with standard deviation $\sigma_{\mu}^{\rm rec}$ (see Table~\ref{table:neutrinos}).

The search identified no \textsc{Antares} neutrino candidates that were temporally coincident with GW150914.

For IceCube, none of the three neutrino candidates temporally coincident with GW150914 were compatible with the GW direction at 90\% CL. Additionally, the reconstructed energy of the neutrino candidates with respect to the expected background does not make them significant. See Fig. \ref{figure:skymap} for the directional relation of GW150914 and the IceCube neutrino candidates detected within the $\pm500$\,s window. This non-detection is consistent with our expectation from a binary black hole merger.

To better understand the probability that the detected neutrino candidates are consistent with background, we briefly consider different aspects of the data separately. %
First, the number of detected neutrino candidates, i.e. 3 and 0 for IceCube and \textsc{Antares}, respectively, is fully consistent with the expected background rate of 4.4 and $\ll1$ for the two detectors, with p-value $1-F_{\rm pois}(N_{\rm observed} \leq 2, N_{\rm expected} = 4.4) = 0.81$, where $F_{\rm pois}$ is the Poisson cumulative distribution function.
Second, for the most significant reconstructed muon energy (Table~\ref{table:neutrinos}), 12.5\% of background events will have greater muon energy. The probability that at least one neutrino candidate, out of 3 detected events, has an energy high enough to make it appear even less background-like, is $1-(1-0.125)^3 \approx 0.33$.
Third, with the GW sky area 90\% CL of $\Omega_{\rm gw} = \PESKYNINTY$, the probability of a background neutrino candidate being directionally coincident is $\Omega_{\rm gw}/\Omega_{\rm all}\approx0.014$. We expect $3\Omega_{\rm gw}/\Omega_{\rm all}$ directionally coincident neutrinos, given 3 temporal coincidences. Therefore, the probability that at least one of the 3 neutrino candidates is directionally coincident with the 90\% CL skymap of GW150914 is $1-(1-0.014)^3 \approx 0.04$.

\subsection{Constraints on the source}

We used the non-detection of coincident neutrino candidates by \textsc{Antares} and IceCube to derive a standard frequentist neutrino spectral fluence upper limit for GW150914 at 90\% CL. Considering no spatially and temporally coincident neutrino candidates, we calculated the source fluence that on average would produce 2.3 detected neutrino candidates. We carried out this analysis as a function of source direction, and independently for \textsc{Antares} and IceCube.

\begin{figure}
\begin{center}
\resizebox{0.49\textwidth}{!}{\includegraphics{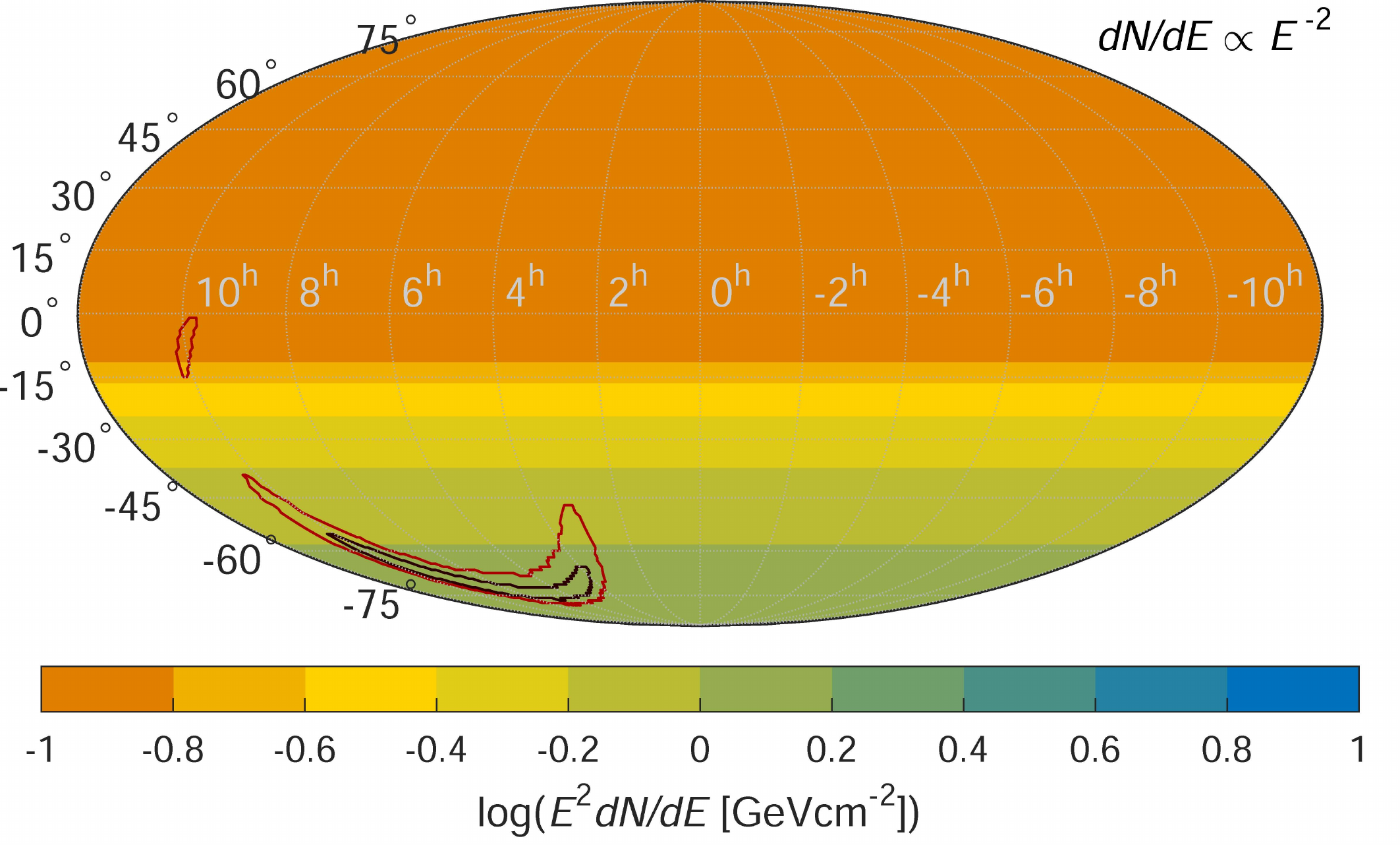}}
\resizebox{0.49\textwidth}{!}{\includegraphics{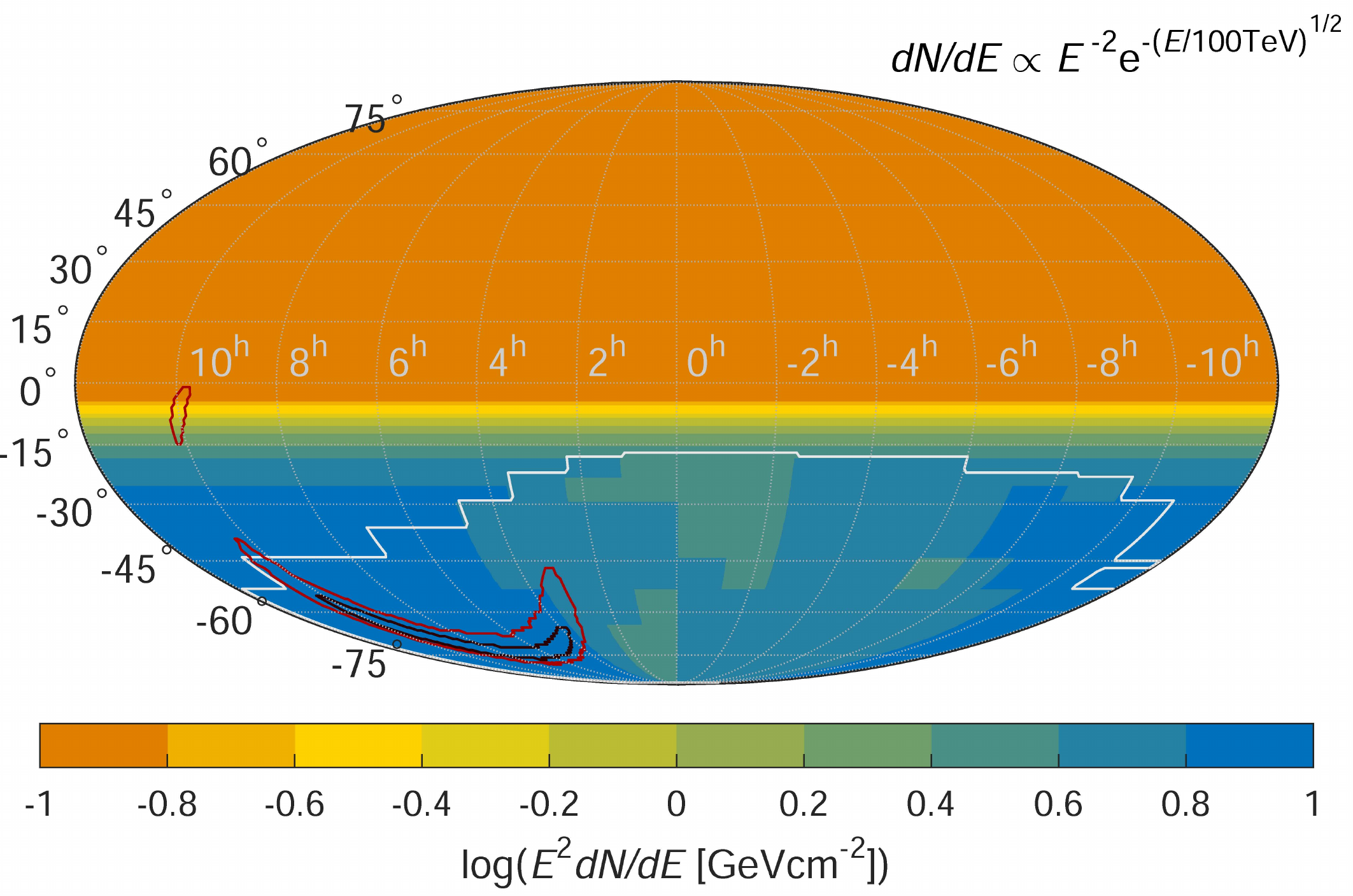}}
\end{center}
\caption{Upper limit on the high-energy neutrino spectral fluence ($\nu_{\mu} + \overline{\nu}_{\mu}$) from GW150914  as a function of source direction, assuming $dN/dE\propto E^{-2}$ (top) and $dN/dE\propto E^{-2}\exp[-\sqrt(E/100\text{TeV})]$ (bottom) neutrino spectra. The region surrounded by a white line shows the part of the sky in which \textsc{Antares} is more sensitive (close to nadir), while on the rest of the sky, IceCube is more sensitive. For comparison, the 50\% CL and 90\% CL contours of the GW sky map are also shown.}
\label{figure:UL}
\end{figure}

The obtained spectral fluence upper limits as a function of source direction are shown in Fig. \ref{figure:UL}. We considered a standard $dN/dE\propto E^{-2}$ source model, as well as a model with a spectral cutoff at high energies: $dN/dE\propto E^{-2}\exp[-\sqrt(E/100\text{TeV})]$. The latter model is expected for sources with exponential cutoff in the primary proton spectrum \cite{2007JPhCS..60..243K}. This is expected for some galactic sources, and is also adopted here for comparison to previous analyses \cite{2015arXiv151102149A}. For each spectral model, the upper limit shown in each direction of the sky is the more stringent limit provided by one or the other detector. We see in Fig. \ref{figure:UL} that the constraint strongly depends on the source direction, and is mostly within $E^{2}dN/dE\sim 10^{-1}-10$\,GeV\,cm$^{-2}$. Furthermore, the upper limits by \textsc{Antares} and IceCube constrain different energy ranges in the region of the sky close to the GW candidate. %(between declinations $-60^{\circ}$ and $-80^{\circ}$).
For an $E^{-2}$ power-law source spectrum, $90\%$ of \textsc{Antares} signal neutrinos are in the energy range from 3\,TeV to 1\,PeV, whereas for IceCube at this southern declination the corresponding energy range is 200\,TeV to 100\,PeV.

To characterize the dependence of neutrino spectral fluence limits on source direction, we calculate these limits separately for the two distinct areas in the 90\% credible region of the GW skymap. For the larger region farther South (hereafter \emph{South region}), we find upper limits $E^2 dN/dE = 1.2^{+0.25}_{-0.36}$\,GeV\,cm$^{-2}$ and $E^2 dN/dE = 7.0^{+3.2}_{-2.0}$\,GeV\,cm$^{-2}$ for our two spectral models without and with a cutoff, respectively. The error bars define the 90\% confidence interval of the upper limit, showing the level of variation within each region. The average values were obtained as geometric averages, which better represent the upper limit values as they are distributed over a wide numerical range. For the smaller region farther North (hereafter \emph{North region}), we find upper limits $E^2 dN/dE = 0.10^{+0.12}_{-0.06}$\,GeV\,cm$^{-2}$ and $E^2 dN/dE = 0.55^{+1.79}_{-0.44}$\,GeV\,cm$^{-2}$. As expected, we see that the limits are much more constraining for the North region, given the stronger limits at the Northern hemisphere due to IceCube's greatly improved sensitivity there. Additionally, we see that the 90\% confidence intervals for the South region, which is much more likely to contain the real source direction than the North region, are fairly small around the average, with the lower and higher limits only differing by about a factor of 2. The upper limits within this area can be considered essentially uniform. We observe a much greater variation in the North region.

To provide a more detailed picture of our constraints on neutrino emission, we additionally calculated neutrino fluence upper limits for different energy bands. For these limits, we assume $dN/dE\propto E^{-2}$ within each energy band. We focus on $\mbox{Dec}=-70^\circ$, which is consistent with the most likely source direction, and also with most of the GW sky area's credible region. For each energy range, we use the limit from the most sensitive detector within that range. The obtained limits are given in Table \ref{table:limitsbands}.

\begin{table}
\begin{tabular}{|r c r|c|}
  \hline
  \multicolumn{3}{|c|}{Energy range} & \,Limit [GeV\,cm$^{-2}$]\, \\ \hline
  100\,GeV & -- & 1\,TeV   & 150                  \\ % 151.9
  1\,TeV   & -- & 10\,TeV  & 18                   \\ % 17.5
  10\,TeV  & -- & 100\,TeV & 5.1                    \\
  100\,TeV & -- & 1\,PeV   & 5.5                    \\
  1\,PeV   & -- & 10\,PeV  & 2.8                    \\
  10\,PeV  & -- & 100\,PeV & 6.5                    \\
  100\,PeV & -- & 1\,EeV   & 28                  \\ % 27.6
  \hline
\end{tabular}
  \caption{Upper limits on neutrino spectral fluence ($\nu_{\mu} + \overline{\nu}_{\mu}$) from GW150914, separately for different spectral ranges, at $\mbox{Dec}=-70^\circ$. We assume $dN/dE\propto E^{-2}$ within each energy band. }
  \label{table:limitsbands}
\end{table}

We now convert our fluence upper limits into a constraint on the total energy emitted in neutrinos by the source. To obtain this constraint, we integrate emission within $[100\,\text{GeV}, 100\,\text{PeV}]$ for each source model. The obtained constraint will vary with respect to source direction as we saw above. It will also depend on the uncertain source distance. To account for these uncertainties, we provide the range of values from the lowest to the highest possible within the 90\% confidence intervals with respect to source direction and the 90\% credible interval with respect to source distance. For simplicity, we treat the estimated source distance and its uncertainty independent of the source direction. We consider both of the distinct sky regions to provide an inclusive range. For our two spectral models, we obtain the following upper limit on the total energy radiated in neutrinos:
\begin{eqnarray}
\mathrm{E}_{\rm\nu,tot}^{\rm ul} &=& 5.4\times 10^{51} \,\text{--}\, 1.3 \times 10^{54}\,\text{erg} \\
\mathrm{E}_{\rm\nu,tot}^{\rm ul \text{(cutoff)}} &=& 6.6\times 10^{51} \,\text{--}\, 3.7 \times 10^{54}\,\text{erg}
\end{eqnarray}
with the first and second lines of the equation corresponding to the spectral models without and with cutoff, respectively. For comparison, the total energy radiated in GWs from the source is $\sim\ERADERGSIMPLE$\,erg. This value can also be compared to high-energy emission expected in some scenarios for accreting stellar-mass black holes. For example, typical GRB isotropic-equivalent energies are $\sim 10^{51}$\,erg for long and $\sim 10^{49}$\,erg for short GRBs \cite{2006RPPh...69.2259M}. The total energy radiated in high-energy neutrinos in the case of GRBs can be comparable \cite{2015arXiv151101396M,2000PhRvL..85.1362B,2013PhRvL.110x1101B,2013PhRvL.111m1102M,2014PhRvD..89d3012M} or in some cases much greater \cite{2001PhRvL..87q1102M,2013PhRvL.111l1102M} than the high-energy electromagnetic emission. There is little reason, however, to expect an associated GRB for a binary black hole merger (see, nevertheless, \cite{2016arXiv160203920C}).

\section{Conclusion}

The results above  represent the first concrete limit on neutrino emission from this GW source type, and the first neutrino follow-up of a significant GW event. With the continued increase of Advanced LIGO-Virgo sensitivities for the next observation periods, and the implied source rate of 2--400\,Gpc$^{-3}$yr$^{-1}$ in the comoving frame based on this first detection \cite{ratescompanion}, we can expect to detect a significant number of GW sources, allowing for stacked neutrino analyses and significantly improved constraints. Similar analyses for the upcoming observation periods of Advanced LIGO-Virgo will be important to provide constraints on or to detect other joint GW and neutrino sources.%, and to aid electromagnetic follow-up surveys by providing accurate source directions.

Joint GW and neutrino searches will also be used to improve the efficiency of electromagnetic follow-up observations over GW-only triggers. Given the significantly more accurate direction reconstruction of neutrinos ($\sim~1$\,deg$^2$ for track events in IceCube \cite{2014PhRvD..89j2004A,2014JInst...9P3009A} and $\sim~0.2$\,deg$^{2}$ in \textsc{Antares} \cite{2014ApJ...786L...5A}) compared to GWs ($\gtrsim 100$\,deg$^2$), a joint event candidate provides a greatly reduced sky area for follow-up observatories \cite{2013ApJ...767..124N}. The delay induced by the event filtering and reconstruction after the recorded trigger time is typically 3--5\,s for \textsc{Antares} \cite{2015arXiv150801180A}, 20--30\,s for IceCube \cite{2015arXiv151005222T}, and $\mathcal{O}(1\,\text{min})$ for LIGO-Virgo, making data available for rapid analyses.

%Since neutrino as well as GW facilities are all-sky detectors, their combined use for triggering follow-up searches are practical.

% ACKNOWLEDGEMENT
%\let\savecleardoublepage\cleardoublepage
%\let\cleardoublepage\clearpage
%\include{GWHEN_acknowledgement}
\begin{acknowledgments}
%\section*{Ackowledgements}
% ANTARES
The authors acknowledge the financial support of the funding agencies:
Centre National de la Recherche Scientifique (CNRS), Commissariat \`a
l'\'ener\-gie atomique et aux \'energies alternatives (CEA),
Commission Europ\'eenne (FEDER fund and Marie Curie Program),
Institut Universitaire de France (IUF),
IdEx program and UnivEarthS Labex program at Sorbonne Paris Cit\'e
(ANR-10-LABX-0023 and ANR-11-IDEX-0005-02), R\'egion
\^Ile-de-France (DIM-ACAV), R\'egion Alsace (contrat CPER), R\'egion
Provence-Alpes-C\^ote d'Azur, D\'e\-par\-tement du Var and Ville de La
Seyne-sur-Mer, France; Bundesministerium f\"ur Bildung und Forschung
(BMBF), Germany; Istituto Nazionale di Fisica Nucleare (INFN), Italy;
Stichting voor Fundamenteel Onderzoek der Materie (FOM), Nederlandse
organisatie voor Wetenschappelijk Onderzoek (NWO), the Netherlands;
Council of the President of the Russian Federation for young
scientists and leading scientific schools supporting grants, Russia;
National Authority for Scientific Research (ANCS), Romania;
Mi\-nis\-te\-rio de Econom\'{\i}a y Competitividad (MINECO), Prometeo
and Grisol\'{\i}a programs of Generalitat Valenciana and MultiDark,
Spain; Agence de  l'Oriental and CNRST, Morocco. We also acknowledge
the technical support of Ifremer, AIM and Foselev Marine for the sea
operation and the CC-IN2P3 for the computing facilities.

%%% ICECUBE %%%%%%%%%%%%%%%%%%%%%%%%%%%%%%%%%%%%%%%%%%%%%%%%%%%%%%%%%%%%%%%%%%%%%%%%%%%%%%%%%%%%%%%%%%%%%%%%%%%
We acknowledge the support from the following agencies:
U.S. National Science Foundation-Office of Polar Programs,
U.S. National Science Foundation-Physics Division,
University of Wisconsin Alumni Research Foundation,
the Grid Laboratory Of Wisconsin (GLOW) grid infrastructure at the University of Wisconsin - Madison, the Open Science Grid (OSG) grid infrastructure;
U.S. Department of Energy, and National Energy Research Scientific Computing Center,
the Louisiana Optical Network Initiative (LONI) grid computing resources;
Natural Sciences and Engineering Research Council of Canada,
WestGrid and Compute/Calcul Canada;
Swedish Research Council,
Swedish Polar Research Secretariat,
Swedish National Infrastructure for Computing (SNIC),
and Knut and Alice Wallenberg Foundation, Sweden;
German Ministry for Education and Research (BMBF),
Deutsche Forschungsgemeinschaft (DFG),
Helmholtz Alliance for Astroparticle Physics (HAP),
Research Department of Plasmas with Complex Interactions (Bochum), Germany;
Fund for Scientific Research (FNRS-FWO),
FWO Odysseus programme,
Flanders Institute to encourage scientific and technological research in industry (IWT),
Belgian Federal Science Policy Office (Belspo);
University of Oxford, United Kingdom;
Marsden Fund, New Zealand;
Australian Research Council;
Japan Society for Promotion of Science (JSPS);
the Swiss National Science Foundation (SNSF), Switzerland;
National Research Foundation of Korea (NRF);
Danish National Research Foundation, Denmark (DNRF)

%%% LIGO VIRGO %%%%%%%%%%%%%%%%%%%%%%%%%%%%%%%%%%%%%%%%%%%%%%%%%%%%%%%%%%%%%%%%%%%%%%%%%%%%%%%%%%%%%%%%%%%%%%%%
The authors gratefully acknowledge the support of the United States
National Science Foundation (NSF) for the construction and operation of the
LIGO Laboratory and Advanced LIGO as well as the Science and Technology Facilities Council (STFC) of the
United Kingdom, the Max-Planck-Society (MPS), and the State of
Niedersachsen/Germany for support of the construction of Advanced LIGO
and construction and operation of the GEO600 detector.
Additional support for Advanced LIGO was provided by the Australian Research Council.
The authors gratefully acknowledge the Italian Istituto Nazionale di Fisica Nucleare (INFN),
the French Centre National de la Recherche Scientifique (CNRS) and
the Foundation for Fundamental Research on Matter supported by the Netherlands Organisation for Scientific Research,
for the construction and operation of the Virgo detector
and the creation and support  of the EGO consortium.
The authors also gratefully acknowledge research support from these agencies as well as by
% the International Science Linkages program of the Commonwealth of Australia,
the Council of Scientific and Industrial Research of India,
Department of Science and Technology, India,
Science \& Engineering Research Board (SERB), India,
Ministry of Human Resource Development, India,
the Spanish Ministerio de Econom\'ia y Competitividad,
the Conselleria d'Economia i Competitivitat and Conselleria d'Educaci\'o, Cultura i Universitats of the Govern de les Illes Balears,
the National Science Centre of Poland,
the European Commission,
the Royal Society,
the Scottish Funding Council,
the Scottish Universities Physics Alliance,
% the National Aeronautics and Space Administration,
the Hungarian Scientific Research Fund (OTKA),
the Lyon Institute of Origins (LIO),
the National Research Foundation of Korea,
Industry Canada and the Province of Ontario through the Ministry of Economic Development and Innovation,
the Natural Science and Engineering Research Council Canada,
Canadian Institute for Advanced Research,
the Brazilian Ministry of Science, Technology, and Innovation,
Russian Foundation for Basic Research,
% the Carnegie Trust,
the Leverhulme Trust,
% the David and Lucile Packard Foundation,
the Research Corporation,
% and the Alfred P. Sloan Foundation.
Ministry of Science and Technology (MOST), Taiwan
and
the Kavli Foundation.
The authors gratefully acknowledge the support of the NSF, STFC, MPS, INFN, CNRS and the
State of Niedersachsen/Germany for provision of computational resources.
This article has LIGO document number LIGO-P1500271.
\end{acknowledgments}

%#################################################################
\bibliographystyle{h-physrev}
%\bibliography{Refs}

  %
  % Restore the author, affiliation and maketitle commands.
  %
  \let\author\myauthor
  \let\affiliation\myaffiliation
  \let\maketitle\mymaketitle
  \title{Authors}
  \pacs{}
    %%%% ANTARES AUTHORLIST %%%%%%%%%%%%%%%%%%%%%%%%%%%%%%%%%%%%%%%%%%%%%%%%%%%%%%%%%
\author{S.~Adri\'an-Mart\'inez}
\affiliation{{Institut d'Investigaci\'o per a la Gesti\'o Integrada de les Zones Costaneres (IGIC) - Universitat Polit\`ecnica de Val\`encia. C/  Paranimf 1 , 46730 Gandia, Spain.}}

\author{A.~Albert}
\affiliation{{GRPHE - Universit\'e de Haute Alsace - Institut universitaire de technologie de Colmar, 34 rue du Grillenbreit BP 50568 - 68008 Colmar, France}}

\author{M.~Andr\'e}
\affiliation{{Technical University of Catalonia, Laboratory of Applied Bioacoustics, Rambla Exposici\'o,08800 Vilanova i la Geltr\'u,Barcelona, Spain}}

\author{G.~Anton}
\affiliation{Erlangen Centre for Astroparticle Physics, Friedrich-Alexander-Universit\"at Erlangen-N\"urnberg, D-91058 Erlangen, Germany}

\author{M.~Ardid}
\affiliation{{Institut d'Investigaci\'o per a la Gesti\'o Integrada de les Zones Costaneres (IGIC) - Universitat Polit\`ecnica de Val\`encia. C/  Paranimf 1 , 46730 Gandia, Spain.}}

\author{J.-J.~Aubert}
\affiliation{{Aix-Marseille Universit\'e, CNRS/IN2P3, CPPM UMR 7346, 13288 Marseille, France}}

\author{T.~Avgitas}
\affiliation{{APC, Universit\'e Paris Diderot, CNRS/IN2P3, CEA/IRFU, Observatoire de Paris, Sorbonne Paris Cit\'e, 75205 Paris, France}}

\author{B.~Baret}
\affiliation{{APC, Universit\'e Paris Diderot, CNRS/IN2P3, CEA/IRFU, Observatoire de Paris, Sorbonne Paris Cit\'e, 75205 Paris, France}}

\author{J.~Barrios-Mart\'{\i}}
\affiliation{{IFIC - Instituto de F\'isica Corpuscular c/ Catedra\'atico Jos\'e Beltr\'an, 2 E-46980 Paterna, Valencia, Spain}}

\author{S.~Basa}
\affiliation{{LAM - Laboratoire d'Astrophysique de Marseille, P\^ole de l'\'Etoile Site de Ch\^ateau-Gombert, rue Fr\'ed\'eric Joliot-Curie 38,  13388 Marseille Cedex 13, France}}

% ...

\collaboration{The {\sc Antares} Collaboration}
\noaffiliation 
    %%% ICECUBE AUTHORLIST %%%%%%%%%%%%%%%%%%%%%%%%%%%%%%%%%%%%%%%%%%%%%%%%%%%%%%%%%%
\affiliation{III. Physikalisches Institut, RWTH Aachen University, D-52056 Aachen, Germany}
\affiliation{University of Adelaide, Adelaide, South Australia 5005, Australia}
\affiliation{Dept.~of Physics and Astronomy, University of Alaska Anchorage, 3211 Providence Dr., Anchorage, AK 99508, USA}
\affiliation{CTSPS, Clark-Atlanta University, Atlanta, GA 30314, USA}
\affiliation{School of Physics and Center for Relativistic Astrophysics, Georgia Institute of Technology, Atlanta, GA 30332, USA}
\affiliation{Dept.~of Physics, Southern University, Baton Rouge, LA 70813, USA}
\affiliation{Dept.~of Physics, University of California, Berkeley, CA 94720, USA}
\affiliation{Lawrence Berkeley National Laboratory, Berkeley, CA 94720, USA}
\affiliation{Institut f\"ur Physik, Humboldt-Universit\"at zu Berlin, D-12489 Berlin, Germany}
\affiliation{Fakult\"at f\"ur Physik \& Astronomie, Ruhr-Universit\"at Bochum, D-44780 Bochum, Germany}
\affiliation{Physikalisches Institut, Universit\"at Bonn, Nussallee 12, D-53115 Bonn, Germany}
\affiliation{Universit\'e Libre de Bruxelles, Science Faculty CP230, B-1050 Brussels, Belgium}
\affiliation{Vrije Universiteit Brussel, Dienst ELEM, B-1050 Brussels, Belgium}
\affiliation{Dept.~of Physics, Massachusetts Institute of Technology, Cambridge, MA 02139, USA}
\affiliation{Dept.~of Physics, Chiba University, Chiba 263-8522, Japan}
\affiliation{Dept.~of Physics and Astronomy, University of Canterbury, Private Bag 4800, Christchurch, New Zealand}
\affiliation{Dept.~of Physics, University of Maryland, College Park, MD 20742, USA}
\affiliation{Dept.~of Physics and Center for Cosmology and Astro-Particle Physics, Ohio State University, Columbus, OH 43210, USA}
\affiliation{Dept.~of Astronomy, Ohio State University, Columbus, OH 43210, USA}
\affiliation{Niels Bohr Institute, University of Copenhagen, DK-2100 Copenhagen, Denmark}
\affiliation{Dept.~of Physics, TU Dortmund University, D-44221 Dortmund, Germany}
\affiliation{Dept.~of Physics and Astronomy, Michigan State University, East Lansing, MI 48824, USA}
\affiliation{Dept.~of Physics, University of Alberta, Edmonton, Alberta, Canada T6G 2E1}
\affiliation{Erlangen Centre for Astroparticle Physics, Friedrich-Alexander-Universit\"at Erlangen-N\"urnberg, D-91058 Erlangen, Germany}
\affiliation{D\'epartement de physique nucl\'eaire et corpusculaire, Universit\'e de Gen\`eve, CH-1211 Gen\`eve, Switzerland}
\affiliation{Dept.~of Physics and Astronomy, University of Gent, B-9000 Gent, Belgium}
\affiliation{Dept.~of Physics and Astronomy, University of California, Irvine, CA 92697, USA}
\affiliation{Dept.~of Physics and Astronomy, University of Kansas, Lawrence, KS 66045, USA}
\affiliation{Dept.~of Astronomy, University of Wisconsin, Madison, WI 53706, USA}
\affiliation{Dept.~of Physics and Wisconsin IceCube Particle Astrophysics Center, University of Wisconsin, Madison, WI 53706, USA}
\affiliation{Institute of Physics, University of Mainz, Staudinger Weg 7, D-55099 Mainz, Germany}
\affiliation{Universit\'e de Mons, 7000 Mons, Belgium}
\affiliation{Technische Universit\"at M\"unchen, D-85748 Garching, Germany}
\affiliation{Bartol Research Institute and Dept.~of Physics and Astronomy, University of Delaware, Newark, DE 19716, USA}
\affiliation{Dept.~of Physics, Yale University, New Haven, CT 06520, USA}
\affiliation{Dept.~of Physics, University of Oxford, 1 Keble Road, Oxford OX1 3NP, UK}
\affiliation{Dept.~of Physics, Drexel University, 3141 Chestnut Street, Philadelphia, PA 19104, USA}
\affiliation{Physics Department, South Dakota School of Mines and Technology, Rapid City, SD 57701, USA}
\affiliation{Dept.~of Physics, University of Wisconsin, River Falls, WI 54022, USA}
\affiliation{Oskar Klein Centre and Dept.~of Physics, Stockholm University, SE-10691 Stockholm, Sweden}
\affiliation{Dept.~of Physics and Astronomy, Stony Brook University, Stony Brook, NY 11794-3800, USA}
\affiliation{Dept.~of Physics, Sungkyunkwan University, Suwon 440-746, Korea}
\affiliation{Dept.~of Physics, University of Toronto, Toronto, Ontario, Canada, M5S 1A7}
\affiliation{Dept.~of Physics and Astronomy, University of Alabama, Tuscaloosa, AL 35487, USA}
\affiliation{Dept.~of Astronomy and Astrophysics, Pennsylvania State University, University Park, PA 16802, USA}
\affiliation{Dept.~of Physics, Pennsylvania State University, University Park, PA 16802, USA}
\affiliation{Dept.~of Physics and Astronomy, University of Rochester, Rochester, NY 14627, USA}
\affiliation{Dept.~of Physics and Astronomy, Uppsala University, Box 516, S-75120 Uppsala, Sweden}
\affiliation{Dept.~of Physics, University of Wuppertal, D-42119 Wuppertal, Germany}
\affiliation{DESY, D-15735 Zeuthen, Germany}

\author{M.~G.~Aartsen}
\affiliation{University of Adelaide, Adelaide, South Australia 5005, Australia}
\author{K.~Abraham}
\affiliation{Technische Universit\"at M\"unchen, D-85748 Garching, Germany}
\author{M.~Ackermann}
\affiliation{DESY, D-15735 Zeuthen, Germany}
\author{J.~Adams}
\affiliation{Dept.~of Physics and Astronomy, University of Canterbury, Private Bag 4800, Christchurch, New Zealand}
\author{J.~A.~Aguilar}
\affiliation{Universit\'e Libre de Bruxelles, Science Faculty CP230, B-1050 Brussels, Belgium}
\author{M.~Ahlers}
\affiliation{Dept.~of Physics and Wisconsin IceCube Particle Astrophysics Center, University of Wisconsin, Madison, WI 53706, USA}
\author{M.~Ahrens}
\affiliation{Oskar Klein Centre and Dept.~of Physics, Stockholm University, SE-10691 Stockholm, Sweden}
\author{D.~Altmann}
\affiliation{Erlangen Centre for Astroparticle Physics, Friedrich-Alexander-Universit\"at Erlangen-N\"urnberg, D-91058 Erlangen, Germany}
\author{T.~Anderson}
\affiliation{Dept.~of Physics, Pennsylvania State University, University Park, PA 16802, USA}
\author{I.~Ansseau}
\affiliation{Universit\'e Libre de Bruxelles, Science Faculty CP230, B-1050 Brussels, Belgium}
% ...

\collaboration{The IceCube Collaboration}
\noaffiliation 
    \author{B.~P.~Abbott}
\affiliation{LIGO, California Institute of Technology, Pasadena, CA 91125, USA}
\author{R.~Abbott}
\affiliation{LIGO, California Institute of Technology, Pasadena, CA 91125, USA}
\author{T.~D.~Abbott}
\affiliation{Louisiana State University, Baton Rouge, LA 70803, USA}
\author{M.~R.~Abernathy}
\affiliation{LIGO, California Institute of Technology, Pasadena, CA 91125, USA}
\author{F.~Acernese}
\affiliation{Universit\`a di Salerno, Fisciano, I-84084 Salerno, Italy}
\affiliation{INFN, Sezione di Napoli, Complesso Universitario di Monte S.Angelo, I-80126 Napoli, Italy}
\author{K.~Ackley}
\affiliation{University of Florida, Gainesville, FL 32611, USA}
\author{C.~Adams}
\affiliation{LIGO Livingston Observatory, Livingston, LA 70754, USA}
\author{T.~Adams}
\affiliation{Laboratoire d'Annecy-le-Vieux de Physique des Particules (LAPP), Universit\'e Savoie Mont Blanc, CNRS/IN2P3, F-74941 Annecy-le-Vieux, France}
\author{P.~Addesso}
\affiliation{Universit\`a di Salerno, Fisciano, I-84084 Salerno, Italy}
\author{R.~X.~Adhikari}
\affiliation{LIGO, California Institute of Technology, Pasadena, CA 91125, USA}
% ...

\collaboration{LIGO Scientific Collaboration and Virgo Collaboration}
\medskip
\author{{$^{\dag}$Deceased, May 2015. $^{\ddag}$Deceased, March 2015.}}
\noaffiliation
\medskip

  \newpage
  \maketitle

\end{document}